\documentclass{sf2a-conf2025}
\usepackage{graphicx}
\usepackage{hyperref}
\usepackage[]{natbib}  
\usepackage{epstopdf}

\def\BibTeX{{\rm B\kern-.05em{\sc i\kern-.025em b}\kern-.08em
    T\kern-.1667em\lower.7ex\hbox{E}\kern-.125emX}}
\bibpunct{(}{)}{;}{a}{}{,}  

\begin{document}

\TitreGlobal{SF2A 2025}


\title{SF2A Environmental Transition Commission:\\ Chosen pieces from the survey 'French A\&A research activities in the face of the environmental crisis, from 2019 to 2024'}

\runningtitle{Review SF2A-2025 S00}

\author{F.~Cantalloube}\address{Univ. Grenoble Alpes, CNRS, IPAG, F-38000 Grenoble, France}
\author{C.~Noûs}\address{Laboratoire Cogitamus, \url{https://www.cogitamus.fr/}}

\author{A.~Jolly}\address{Univ Paris Est Creteil et Universite Paris Cite, CNRS, LISA, F-94010 Creteil, France}
\author{J.~Milli$^1$}
\author{J.-F.~Gonzalez}\address{Universite Claude Bernard Lyon 1, CRAL UMR5574, ENS de Lyon, CNRS, Villeurbanne 69622, France}
\author{F.~Laurent$^4$}




\maketitle

\begin{abstract}
In 2025, the French Society for Astronomy \& Astrophysics (SF2A), gave the environmental transition commission the opportunity to share their considerations during a plenary session at the annual SF2A conference. This year, the presentation focused on some of the main results obtained from the survey entitled \textit{French astronomy and astrophysics research activities in the face of the environmental crisis, from 2019 to 2024}. The survey was initiated in 2019 by the group \emph{Environnement-Transition} (coordinated by P.~Martin) at IRAP, whose results were presented during the SF2A annual conference 2019 in Nice. The survey was updated in 2024 by the CNRS INSU-AA prospective working group \emph{Climate and ecological challenge} (coordinated by S.~Bontemp). The SF2A environmental transition commission took on the survey to the French  institutes, sorted the answers and extracted the preliminary results. The full results will be published at the end of 2025 in the final CNRS INSU-AA 2024 prospective document. This publication presents a selection of pieces from the full survey, along with a few of the main discussions it triggers. 
\end{abstract}



\section{Introduction}
In 2025, the annual conference of the French Society of Astronomy and Astrophysics (SF2A), a non-profit organization gathering more than 500 professional researchers working in astronomy and astrophysics, took place in Toulouse, France. During the conference in early July, the temperature rose to $40^\circ C$ ($+16^\circ C$ compared to the daily peak average during pre-industrial era according to reanalysis data\footnote{See for instance the Berkeley database \url{https://berkeleyearth.org/} or website focusing on big cities \url{https://www.city-facts.com/toulouse/weather?utm_source=chatgpt.com}}). Working under such conditions is becoming increasingly unbearable, even for research occupations, as demonstrated by a number of recent publications showing its impact on cognitive performance \citep{yin2024}. Additionally, our daily work tasks heavily rely on numerical tools, such as personal laptops and data centers, whose functioning is vulnerable to high temperatures \citep{Manousakis2016}. In academia, work performance is intertwined with mental health \cite{marais2020,nicholls2022}, which is somewhat unfortunate given that heatwaves increase the risk of mental and mood disorders \cite{liu2021}. 

Toulouse is the $4^{th}$ largest city in France in terms of number of inhabitants (more than 500,000 within the city limits and 1,500,000 outside), as well as in terms of economic weight using the GDP as a metric. The local economy is essentially driven by the aviation and aerospace industry, with notably the main establishment of the French national space agency, CNES, created in 1961 (approximately 1,700 employees) and the parent company of Airbus created in 1970 (approximately 13,000 employees). 
The Toulouse-Blagnac airport is the $7^{th}$ most used in France with 7,844,953 passengers in 2024, including 34\% traveling to/from Paris-Orly airport (the most frequented airline in Europe). 
This high traffic is not only due to a lack of public investment for the construction of a high-speed train line that would connect Toulouse to Paris, but the airport's history itself is marred by controversies and scandals around its privatisation in 2015 \citep{coursdescomptes2018}. 
The Toulouse industrial activities in the past engendered an important soil contamination, particularly with lead, mercury, and arsenic. Contemporary industrial activities also contribute to pollution. In September 2001, the traumatic explosion of the AZF, a factory producing nitrogen fertilizer mainly intended for intensive agriculture had been the source of significant and severe human health and material damages. Another symbolic example of deadly policy is the on-going A69 highway project: 2,000 scientists have publicly denounced it as both a social and ecological absurdity, outweighing the claimed regional benefit. This project that goes against the public interest, exists solely because private interests want to see it through to completion (with conflicts of interest notably involving the Pierre Fabre group). The hunger strikes against this project were ignored and the repression was enormous and disproportionate. For instance, on the $22^{nd}$ of October 2023, a scientific conference given by our colleagues from Atécopol\footnote{\url{https://atecopol.hypotheses.org/9021}} (whose 2025 conferences took place in Toulouse at the same time as the SF2A conference), discussing about the reasons for rejecting this project was violently interrupted by law enforcement officers (using tear gas grenades on a private land), while vulnerable individuals, such as children and elderly people, were in attendance \citep{A692024}. 

In the previous paragraph, the three examples given are to highlight the intrinsic link between environmental degradations as a whole and the political decisions taken in three different cases, at various levels and timeframes. 
While the latest annual report published by the \emph{Haut Conseil pour le Climat} \citep{hce2023} states that France is not prepared for current and future climate change, the recent governments are sharply cutting the budget allocated to ecology (one of the most severely cut categories). In addition, recent years have seen a sharp increase in legal proceedings and criminalization of citizens and activists who are trying to protect the biosphere from ultraliberal greed. 
Meanwhile, it has been confirmed that the seventh of the nine planetary boundaries has been crossed \citep{kitzmann2025}... 
In this context, the objective of the survey partial results presented hereafter is to evaluate and discuss how the French astronomy and astrophysics research community views the commitments it could and should make in response to the rapid and drastic environmental changes affecting the Earth, accounting to the current social context and specificity of our community.




\section{Survey results}
The purpose of the survey is to gauge the position of the French astronomy and astrophysics research community regarding environmental emergencies and the necessary transition for its research activities to align with environmental, and therefore social, ethics. The updated 2024 version of the survey is aimed at all astronomy and astrophysics research staff working in France, including technical, administrative and non-permanent staff. 
We received a total of $388$ valid answers for the 2024 survey\footnote{\url{https://journees.sf2a.eu/wp-content/uploads/2025/07/S00_Cantalloube_online.pdf}} and about $500$ for the 2019 survey\footnote{\url{https://sf2a.eu/semaine-sf2a/2019/presentations/S00/Martin_S00.pdf}}. This represents about one third of the French professional researchers in astronomy and astrophysics.

The survey is organized in four blocks: (1) the identity of the respondent in 5 questions, (2) the awareness level of the respondent in 7 questions, (3) eighteen affirmations to be compared to 2019 using five gradients of answer (\emph{strongly agree}, \emph{rather agree}, \emph{moderately agree}, \emph{rather disagree}, \emph{strongly disagree}), and one \emph{without opinion} option, (4) four optional open-ended questions, which allowed respondents to express themselves freely. This survey has a number of known limitations and is intended to provide an overview of the community's major opinions, particularly regarding the perception on the topics discussed and their evolution over the five last years. The questions raised topics about (i) research ethics and culture, (ii) raising awareness and education, (iii) the concept of exemplarity, (iv) individual vs. institutional actions, and (v) career evaluation. 
The following present a subjective overview analysis of the survey results. It is purposely not complete, focuses only on some key points, and it does not take into account potential correlations, as a more detailed analysis is planned for a subsequent publication. 

\paragraph{Identity of respondents}
Among the $388$ valid answers, 97.4\% of respondent work in France, 29\% of women and 64\% of men, which is representative of the French astronomy and astrophysics research community. The vast majority of respondent are permanent researchers ($57.6\%$), followed by research support staff ($27.8\%$) and non-permanent researchers ($14.6\%$). 
Compared to 2019, we received fewer than half of the responses from people under 30 years old (the demographics of respondents peak among those aged 40 to 45). Among the respondents, $20\%$ remembered responding to the 2019 survey, while $46\%$ did not remember.

\paragraph{Awareness}
To start with awareness, we asked the respondents if they had read the CNRS ethics committee special report produced in 2022 (Avis $n^\circ2022-43$) entitled \emph{Integrating environmental issues into research conduct – An ethical responsibility}: $48\%$ stated to have read it, while $15\%$ do not remember. 
Four questions were about the evolution between 2019 and 2024: (1) whether respondents felt they had become more aware of environmental issues in the workplace, (2) whether they had become better informed about it, (3) whether they feel they had been more active in changing their habits in their personal life, and (4) in their professional life. For these four questions, between 70-80\% of respondents answered positively (\emph{strongly agree} and \emph{somewhat agree}), while less than 3\% answered negatively (\emph{not at all}, which may also be due to people being aware and changing their lifestyle before 2019). 
At last we asked if respondent are aware of any active initiatives in the field of environmental considerations in astronomy and astrophysics research, more than 15 initiative where mentioned. The most mentioned one being the \emph{Labos1point5} collective, then the SF2A special commission and the special working group of the CNRS INSU-AA prospective. Other mentions were for groups within institutes, awareness tools, training programs, and publications (scientific or in the mainstream press).

\paragraph{From 2019 to 2024, what has changed}
Respondents feeling that there is a general evolution in research practices went from $38\%$ to $50\%$. 
About limiting business trips, there was a considerable leap, in terms of encouragement by the management team (from $5$ to $32\%$) and in terms of using train instead of plane when possible (from $8$ to $50\%$). About assessing the carbon footprint of institutes, there was also a considerable shift as more respondent considered having sufficient tools (from $3$ to $26\%$) and more respondent have somehow participated to the footprint assessment (from $21$ to $59\%$). 

\paragraph{From 2019 to 2024, what has not changed}
About 35\% of respondents feel that limiting their business trips will affect their career progression. 
About 52\% of respondents stated that efforts to reduce their environmental impact should be considered in career evaluation. 

\paragraph{From 2019 to 2024, what remains positive}
\textit{For the following propositions, we aggregated the fraction of respondent who answered \emph{strongly agree} and \emph{rather agree}. Figures are shown for the years 2019 to 2024.}\\ 
(p1) Public research has the duty to lead by example ($91.6$ to $75.6\%$), \\
(p2) it must define a multi-year plan with clear and quantified objectives ($83.6$ to $81.8\%$), \\
(p3) the environmental impact of infrastructures must be a major criterion in decision-making ($64$ to $56\%$),\\
(p4) environmental issues must play an important role in the community during major events that shape the community and scientific discussions ($68$ to $67\%$).

\paragraph{From 2019 to 2024, what remains negative}
\textit{For the following propositions, we aggregated the fraction of respondent who answered \emph{rather disagree} and \emph{strongly disagree}, and the \emph{moderately agree} after the $+$ sign.} \\
(n1) Public research changed its practices in response to climate change warnings ($69.6+17\%$ to $27.7\%+46\%$),\\
(n2) public research agencies and universities implemented policies that are commensurate with the challenges of ecological transition ($78.6+14.4\%$ to $54\%+33\%$).

\paragraph{Other propositions 2024, what remains divided}
\textit{In the following propositions, each of the 6 gradient of answers received between $10$ and $20\%$ equally}. \\
(d1) Reducing business travel by a factor of two or three would affect one's career progression,\\
(d2) reducing the number of instrumental projects would improve one's quality of life at work,\\
(d3) reducing our environmental impact would make us less competitive internationally.

\paragraph{Other proposition 2024, what is rather in agreement}
\textit{In the following propositions, about $60\%$ of respondents have answered \emph{strongly agree} or \emph{rather agree}}.\\
(a1) It is possible to conduct high-quality, competitive research with minimal environmental impact,\\
(a2) given that 70\% of greenhouse gas emissions are due to research infrastructures, it is conceivable to reduce the number of instrumental projects, \\
(a3) with the budget redistributed to benefit Research \& Development activities ($52.7\%$),\\
(a4) with the budget redistributed to benefit an increase in Human Resources ($62.8\%$).

\paragraph{Other proposition 2024, the clear consensus}
\textit{In the following propositions, about $90\%$ of respondents have answered \emph{strongly agree} or \emph{rather agree}}.\\
(c1) Research ethics must incorporate environmental issues,\\
(c2) environmental issues represent a major challenge for the community,\\
(c3) they will require profound changes in the way we do and manage research, \\
(c4) institutions should strongly encourage each of us to reduce their greenhouse gas (GHG) emissions.



\section{Discussion of the survey results}
As of today, and essentially starting back in 2019, many actions have been undertaken to initiate discussions about environmental issues within the academic community, especially in the fields of astronomy and astrophysics.  
First, by raising awareness among our colleagues and making the topic visible and meaningful, in particular through various communication tools and collectives gathering and sharing of knowledge.  
Then the community quantified the situation around three areas: (1) the impact of climate change and environmental crises on astronomical research activities, (2) the impact of astronomical research activities on the environment through infrastructures, numerical tools, business travel, and the functioning of institutes, and (3) gauging perceptions and visions of environmental crises within our community through surveys. Now, as of 2025, we have all the figures and qualitative studies necessary to understand the problem, its causes, and its consequences.

\subsection{Decision on an investment project}
One of the proposition raised in the survey was the following: \emph{If we don't make the ecological transition now, it will be imposed on us later, and it won't be the way we want}. To that, $42.3\%$ of respondent \emph{strongly agree}, $27.3\%$ \emph{rather agree}, $26.6\%$ \emph{moderately agree}, $7.3\%$ \emph{rather disagree}, and $4.6\%$ \emph{strongly disagree}. 

If the community now wishes to take environmental impacts into account, several considerations must be taken into account when deciding whether to launch or continue a project, including the following: (i) Will the planned investment be able to function in a changing climate, where temperatures are rising and its consequences? (ii) Will this investment preserve the ecosystem in which it is installed, or will it intensify habitat destruction? (iii) Will this investment reduce GHG emissions in line with climate change mitigation strategies? (iv) Will this investment respect the local population without depleting its resources and without distorting the distribution of resources in favor of a specific segment of the population? (iv) Will this investment avoid any geopolitical interaction that would make them accomplice of governments violating human rights?

In addition, one aspect that must be taken seriously, and that was emphasized during the CNRS INSU-AA prospective in 2024, is that the amount of astronomical data obtained today is beginning to exceed by far the necessary human resources for analysis and interpretation. At last, in the context of astronomical projects, the decisions taken now are usually long-term projects that demand high commitment from researchers and funding agencies. Considering the longer timelines of astronomical projects, most of these projects will concern the careers of today's junior researchers that are unfortunately not part of the decision making process. Aware of that, the community has the duty to ensure that research in astronomy is resilient in the face of the future.

\subsection{Inertia in the astronomy and astrophysics research community}
From the results of the survey exposed in the previous section, there is a clear consensus that the astronomy and astrophysics French research community recognizes the importance of environmental issues and the need for structural changes in the way we conduct research (c1-4). 

However, in practice, it seems that most researchers are reluctant to express their opinions in the context of their work and refrain from gathering in groups to put forward their views. This phenomenon is usually referred to as \emph{the spiral of silence}, where individuals remain silent about their environmental concerns if they believe their views are in the minority, which reinforces the perception that few people care, further discouraging open discussion. This self-reinforcing cycle can hinder public awareness and collective action on environmental issues. This would explain why there is a gap between the desire for change and actual inaction.

Another study by \cite{dupont2025} argues that despite the growing scientific understanding about biodiversity loss and climate change, many researchers remain insufficiently engaged in action. They identified four key barriers: institutional constraints, norms of scientific objectivity, lack of training in engagement, and fear of professional repercussions. These obstacles decouple awareness from concrete responses and call for collective reflection and reorientation of scientific roles. Other studies, such as the one by \cite{dablander2024}, also attempt to explain the many obstacles to action and commitment.

In such context, we may expect senior permanent researchers to commit more than junior researchers, who are tied to the academic system and must "play by the rules" to aspire to a long-term job in the field. However, we observe the opposite effect. Early-career researchers feel that, given the scale of the environmental and social crisis, they have nothing to lose. Permanent staff, on the other hand, might be afraid to lose their positions and privileges, which can be perceived as a form of violence.

Changing the way we conduct research requires breaking down the systemic barriers that place the burden of problem and its solutions on individuals, breaking social ties and causing conflicts rather than connections. Given the constraints and limitations in the astronomy research community, it is necessary to set up working groups based on collective intelligence in order to move forward in a healthy, inclusive, and optimal manner.

\subsection{The pitfalls of moralism and legitimacy}
In Europe, and particularly in France, the scientific education emphasizes  building mathematical models and quantifying everything before making any judgments about importance. This rigid, Cartesian approach to science often overlooks the value of qualitative insights and instinct, which are also deeply informed by knowledge. As a result, our profession is marked by a strong fear of error and an almost reverential commitment to scientific certainty. This fear of being political or stepping outside one's expertise makes many researchers reluctant to speak openly about the real-world implications of data and findings.

Thus, by this education, scientists are told to distinguish between facts and personal opinions. However, the right to make mistakes, to be imperfect, and not to know everything about all subjects related, or not, to our research topics should not prevent us from deciding to act and commit ourselves to the common good. Moralism, the tendency to judge actions or people strictly through a moral code, can lead to rigid thinking and may dismiss complexity, context, or differing values. The current social and environmental crises should encourage researchers to step outside their comfort zone and consider that their role also involves tackling the unknown in interdisciplinary and transdisciplinary fields, and making their voices heard to highlight possible ways forward. It should invite to break out of this pseudo-neutrality that consists of hiding behind numbers.

About legitimacy, the work by the historian of science N.~Oreskes unveiled that three american physicists, the solid-state physicist F.~Steitz, the nuclear physicist W.~Nierenber and the astrophysicist R.~Jastrow, played an essential role in climate change denial, `by challenging the scientific evidence of anthropogenic causes of the ozone hole and global warming' \citep{oreskes2010}. To be noted that they also `spoke strongly against communism and in favour of free enterprise' \citep{oreskes2010}. On the other hand, in 1985, astrophysicists C.~Sagan testified before Congress on climate change reality and its anthropogenic origins \footnote{\url{https://www.youtube.com/watch?v=Wp-WiNXH6hI}}. If the researchers who approved proposal (c1-4) in the survey were to come together and take a stand, this would potentially leave less room for detractors in the future.



\subsection{Towards renunciation?}
In 2021, the European Organization for Nuclear Research (CERN) announced the project to extend the $27~\mathrm{km}$-long Large Hadron Collider (LHC) to a $90~\mathrm{km}$-long Future Circular Collider (FCC) in order to unveil the deepest secrets of particle physics. According to the project's promoters, this would be the only way to prove the existence of certain particles. Of course, it would come at an enormous environmental cost and require an unprecedented amount of energy to operate. Considering the current social, economical and environmental context, a group of scientists\footnote{\url{https://scientifiquesenrebellion.fr/textes/positionnements/tribune-cern/}} questioned the desirability of developing such scientific research infrastructure and called for the project to be abandoned for the sake of environmental protection. 

The same dilemma arises in astronomy. The so-called `fundamental questions' of modern astrophysics drive large-scale missions, cutting-edge observatories, and complex simulations across the field. But in a world where a single species pursues this knowledge on a finite planet, we must ask: at what cost are these answers sought, for whose benefit, and to serve what purpose? At what pace should it be pursued? And ultimately, do we truly need these answers at all?



\begin{figure}[ht!]
 \centering
 \includegraphics[width=\textwidth,clip]{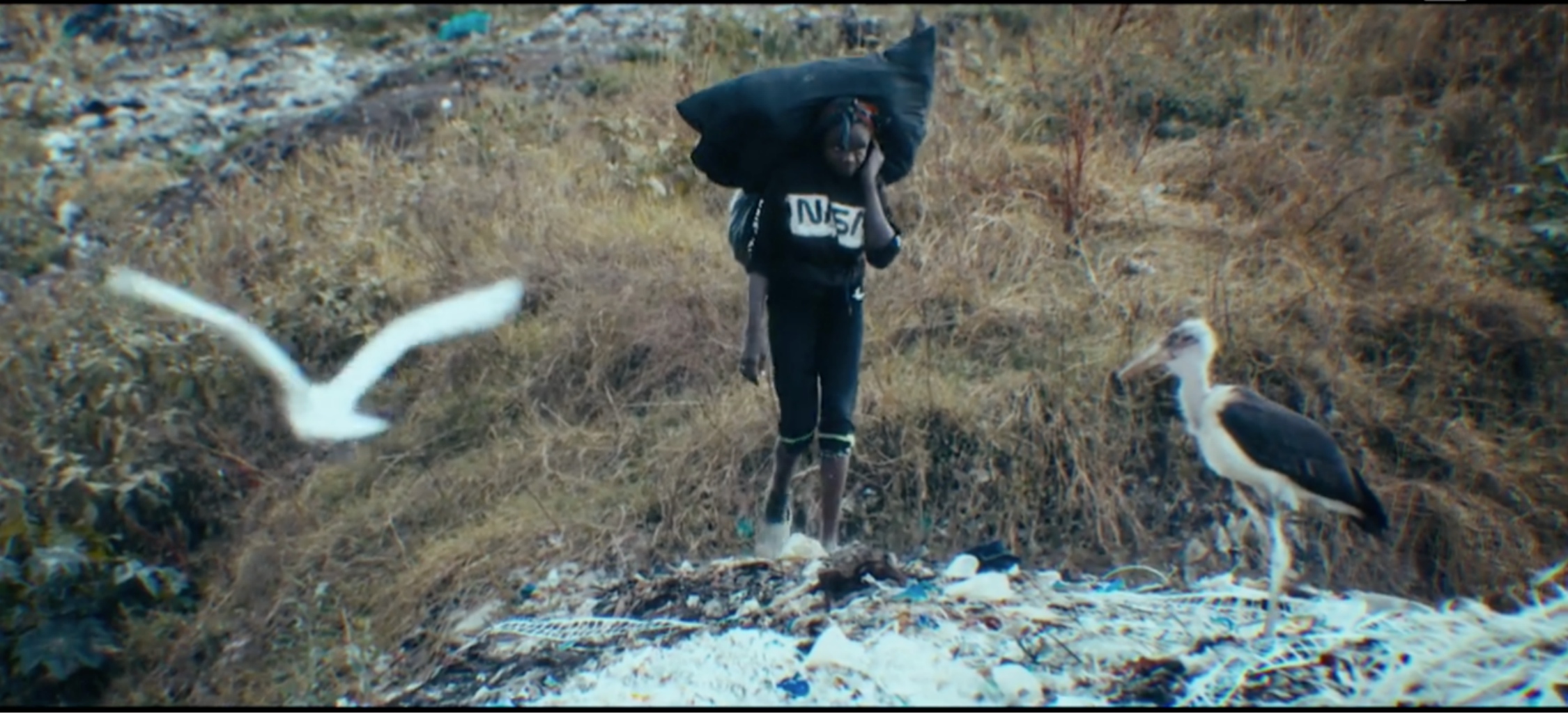}    
  \caption{Landfill in Kenya, made of waste from the USA flown by aircraft. Screenshot from the documentary \emph{Les sacrifiés de l'IA} (In the Belly of AI), directed by H.~Poulain and broadcast by FranceTV in 2025.}
\end{figure}

\section{Conclusions}
Faced with environmental catastrophe, it is more urgent than ever to take a collective stand to protect the public interest against a handful of privileged individuals who want to impose their ultra-liberal and authoritarian ideology. 
As professional astronomers, we must also recognize that we are part of a dominant society. A dominant society whose wealth is based on the exploitation, enslavement, and control of living beings and their substrate. A dominant society that ranks values to justify, establish, and perpetuate its system of domination. 
A dominant society that causes a fantasized amalgamation between progress and technological services with truncated or even inaccurate metrics.

After the second world-war, 90\% of the city of Warsaw was destroyed by subsequent bombing. The pictures of the city taken at the time are terrifying. Guides and books describe the city as being like \emph{the surface of the Moon}. As of today, a major certainty is that life is on Earth and outside of its protective atmosphere, human beings depend on technology to avoid death. As of today, and certainly for many years to come, all of humanity does not have the physical, psychological and technical capacities to settle on a different planet. Turning our gaze, our hopes, and our investments toward space exploration, such as moon settling, is almost criminal in the face of the current destruction of Earth's biosphere. Destruction caused by harmful human activities, motivated by greed rather than the interests of all forms of life, human and non-human. As long as the planet Earth has a finite size, it is criminal to claim that technological developments might save humans, while so far, it has been only worsening the problem and digging inequalities. 

As a trusted community, it seems reasonable to devote some time to working together toward a better future for everyone, including all forms of life, by protecting as much as possible the ecosystem in which they can thrive, including our own. Once again, we cannot engage in astronomy and continue to dream of other worlds while our own reality is being destroyed. Especially when this destruction is caused by a minority, albeit one that is powerful and influential in the mechanisms of today's world, in which most humans seem trapped. Hopefully, there are different stories to be told in the future.

\begin{acknowledgements}
We thank the SF2A board members for supporting the survey initiative and distributing the links to the SF2A members. F.~C. would like to warmly thank the members of the SF2A board for giving her carte blanche during the plenary session and for trusting her to speak freely. The text summarizes the reflections that emerged from discussions with many colleagues from various fields.
\end{acknowledgements}

\bibliographystyle{aa}  
\bibliography{Cantalloube_S00} 

\begin{thebibliography}{12}
\expandafter\ifx\csname natexlab\endcsname\relax\def\natexlab#1{#1}\fi

\bibitem[{{Cour des Comptes}(2018)}]{coursdescomptes2018}
{Cour des Comptes}. 2018, Le processus de privatisation des aéroports de
  Toulouse, Lyon et Nice, Communication à la commission des finances, de
  l’économie générale et du contrôle budgétaire de l’Assemblée
  nationale

\bibitem[{Dablander {et~al.}(2024)Dablander, Sachisthal, \&
  Haslbeck}]{dablander2024}
Dablander, F., Sachisthal, M.~S., \& Haslbeck, J.~M. 2024, npj Climate Action,
  3, 105

\bibitem[{Dupont {et~al.}(2025)Dupont, Jacob, \& Philippe}]{dupont2025}
Dupont, L., Jacob, S., \& Philippe, H. 2025, Nature Ecology \& Evolution, 9, 23

\bibitem[{{HCE}(2023)}]{hce2023}
{HCE}, H. C. p. l.~C. 2023, Rapport annuel 2023

\bibitem[{Kitzmann {et~al.}(2025)Kitzmann, Caesar, Sakschewski, Rockstr{\"o}m,
  Andersen, Bechthold, Bergfeld, Beusen, Billing, Bodirsky,
  {et~al.}}]{kitzmann2025}
Kitzmann, N., Caesar, L., Sakschewski, B., {et~al.} 2025

\bibitem[{Liu {et~al.}(2021)Liu, Varghese, Hansen, Xiang, Zhang, Dear, Gourley,
  Driscoll, Morgan, Capon, {et~al.}}]{liu2021}
Liu, J., Varghese, B.~M., Hansen, A., {et~al.} 2021, Environment international,
  153, 106533

\bibitem[{Manousakis {et~al.}(2016)Manousakis, Sankar, McKnight, Nguyen, \&
  Bianchini}]{Manousakis2016}
Manousakis, I., Sankar, S., McKnight, G., Nguyen, T.~D., \& Bianchini, R. 2016,
  in 14th USENIX Conference on File and Storage Technologies (FAST 16) (Santa
  Clara, CA: USENIX Association), 53--65

\bibitem[{Marais {et~al.}(2020)Marais, Lantheaume, Fiault, \&
  Shankland}]{marais2020}
Marais, G.~A., Lantheaume, S., Fiault, R., \& Shankland, R. 2020, European
  Journal of Investigation in Health, Psychology and Education, 10, 1035

\bibitem[{Nicholls {et~al.}(2022)Nicholls, Nicholls, Tekin, Lamb, \&
  Billings}]{nicholls2022}
Nicholls, H., Nicholls, M., Tekin, S., Lamb, D., \& Billings, J. 2022, PloS
  one, 17, e0268890

\bibitem[{Oreskes \& Conway(2010)}]{oreskes2010}
Oreskes, N. \& Conway, E.~M. 2010, Nature, 465, 686

\bibitem[{Welzer-Lang \& Toustou(2024)}]{A692024}
Welzer-Lang, D. \& Toustou, M. 2024, PhD thesis, Lisst-cers Cnrs, Ligue des
  droits de l'Homme

\bibitem[{Yin {et~al.}(2024)Yin, Fang, Liu, Guo, Ma, \& Di}]{yin2024}
Yin, B., Fang, W., Liu, L., {et~al.} 2024, Ecotoxicology and environmental
  safety, 275, 116238

\end{thebibliography}

%
\end{document}